\documentclass[a4paper,10pt]{article}
\usepackage[utf8]{inputenc}
\usepackage{amssymb,amsmath}
\usepackage{authblk}
\usepackage[left=0.5in, right=0.5in, top=1in, bottom=1in]{geometry}
\usepackage{hyperref}

\title{Thermodynamics of ideal Bose gas under generic power law potential in $d$-dimension}

\author{Mir Mehedi Faruk\\
Department of Theoretical Physics, University of Dhaka, Dhaka-1000\\
\href{mailto:me@somewhere.com}{Email: muturza3.1416@gmail.com, mehedi.faruk.mir@cern.ch} 
 }

\begin{document}
\maketitle
 
 \begin{abstract}
Thermodynamic properties of ideal Bose gas trapped in an external generic power law potential
are investigated systematically from the grand thermodynamic potential in $d$ dimensional space. The most general conditions
for Bose-Einstein condensate and the discontinuous conditions of heat capacity at the critical temperature
in presence of generic power law potential are presented in this manuscript. The dependence of the physical quantities on external potential,
particle characteristics and space dimensionality are discussed.
The more general results
obtained in this paper presents a unified illustration of Bose-Einstein condensation of ideal Bose systems 
as they reduces to the expressions and conclusions available in the literature with appropiate choice of power law exponent.
\end{abstract}

\section{Introduction}
Many attempts have been made to understand the phenomenon
of Bose-Einstein condensation (BEC), after the demonstration by Einstein that there is a possibility of condensation
of free bosons \cite{bose,einstein}. Since then the bulk behavior  of bose gas are investigated  by
many authors\cite{may,robinson, landau, pathria, huang, C. J. PETHICK,ziff,fujita,beckmann,beckmann2}. An increasing 
attraction towards this subject is observed, after the achievement to create BEC in magnetically trapped alkali gases
\cite{Bradley,anderson,davis}.  As a result, different studies analyzing effects of external 
potentials\cite{sala, dal, toroidal, jellal, chen},
relativistic effect\cite{beckmann,beckmann2,grether}, space dimensionality\cite{sala, jellal, acharyya}, 
criticality near the transition point\cite{marek1,gambassi}, 
casimir effect\cite{marek1, bhuiyan, biswas}  of the bosonic system have been performed. Moreover, the recent experimental
observation\cite{nature} of condensation of massless bosons makes the research area even more challenging. \\\\
In a real system, of course  interaction between particles do exist. But taking it into account makes the problem difficult
to solve analytically. Interactions between bosons and their effect on critical temperature $T_c$ and 
fraction of condensation $\frac{N_0}{N}$ are understood for experimental situation, which are few percent or less when 
the density of 
gas is low\cite{ensher}. Thus it is well approximated that the Bose gas of low density can be treated as in ideal bose gas.
Neveretheless, to understand the effect of interactions and to retain the essential physics, we approximately  represent 
the real system by non interacting particles in the presence of an external potential such as harmonic potential 
\cite{dal}, 
toroidal potential\cite{toroidal},
power law potential\cite{sala, chen},
mean field potential\cite{bhuiyan}. The constrained role of external potential for atomic gases
do change the performence of gases. Thus trapped atomic gases provide the opportunity to manipluate 
the  quantum statistical effects. An excellent study by Luca Salanasich \cite{sala}on  Bose gas confined in a trap, described by a 
symmetric (isotropic)
power law potential 
($U=Ar^n$) in d-dimensional space presents a very important relation whether the system will exhibit condensation
or not. Salanasich draw a conclusion from the argument of number of particles that, BEC exists if $\frac{d}{2}+\frac{d}{n}>1$, 
where $n$ is the exponent of symmetric power
law potential. Also, from this relation 
one can say BEC is possible in trapped Bose gas under this type of potential
in $d=2$ with appropiate choice of $n$, which was not the case for ideal free Bose gas. So, it would be interesting to investigate the relation
when the potential is not symmetric in general. 
A point to note, BEC for free bosons at $d=3$ is a first order phase transition\cite{pathria,huang,ziff}. 
But the order of phase transition for the trapped bosons in  arbitrary  is yet to be examined.\\ \\
Lot of efforts have been made in order to understand dimensional dependence of phase 
transition\cite{may, ziff, beckmann,beckmann2, sala, jellal, acharyya, kim}.
Thermodynamic as well as other properties are calculated in
d-dimension for bosons  as well as Fermions\cite{acharyya2, acharyya3, acharyya4} and 
interesting conclusions such as  equivalence
relation between the Bose and Fermi gases in two dimensions is established\cite{lee}. 
In this report we intend to calculate all 
the thermodynamic properties of ideal Bose system in $d$ dimension, under generic
power law potential
(not essentially symmetric). In order to do this,
we have first calculated the grand thermodynamic potential.
And then from grand potential, we have calculated 
the thermodynamic quantities such as internal energy $E$, entropy $S$, pressure $P$, number of
particle $N$, 
helmholtz free energy $A$, isothermal compressibilty $\kappa$, specific heat at constant volume $C_V$ and pressure $C_P$,
their ratio $\gamma=\frac{C_P}{C_V}$.
From the thermodynamic quantities we have derived important results 
regarding the Bose system   such as condensate fraction, specific heat jump, latent heat of condensation, critical temperature, equation of state
as well as  general criteria for existance of BEC for trapped boson under generic power law potential. 
Beside this, from the Clapaeyron equation it has been shown that, like the free ideal bosons in d=3\cite{huang,ziff},
BEC for trapped bosons is also a first order phase transition in any dimension. 
A point to note that,  in the hamiltonian
instead of $\frac{p^2}{2m}$ type kinetic part, we have took $ap^s$, where $p$ is momentum and $a$
is constant. And thus making the important conclusions in more generalized way and obviously the results
obtained here provide a unified description of  behavior of trapped Bose gases, from which many
results in the literature can be derived.\\\\
The report is organized in the following way. The density of states and grand potential are calculated in section 2. Section 3 is devoted to investigate the thermodynamic quatities and to check the order of phase transition. Results and discussions are presented in section 4. The report is concluded in section 5. 

\section{Density of States  and grand potential of bose gas under generic power law potential in d dimension}
Let us consider the ideal  Bose gas in a confining external potential in a d-dimensional space with energy spectrum, 

\begin{eqnarray}
\epsilon (p,x_i)= bp^l + \sum_{i=1} ^d c_i |\frac{x_i}{a_i}|^{n_i}
\end{eqnarray}
Where, $b,$ $l,$ $a_i$, $c_i$, $n_i$  are all postive constants, $p$ is the momentum 
and $x_i$ is the  $i$ th component of coordinate of a particle. Here, $c_i$, $a_i$ and $n_i$ determines the depth 
and confinement power of
the potential. With $l=2$, $b=\frac{1}{2m}$ one can get the energy spectrum  of  the hamiltonian used in 
the literatures \cite{pathria,huang,ziff,sala}.
For the free system all $n_i\longrightarrow \infty$. \\\\
Density of states can be obtained from the following formula,
\begin{eqnarray}
 \rho(\epsilon)&=& \int \int \frac{d^d r d^d p}{(2 \pi \hslash)^d} \delta(\epsilon - \epsilon (p,r))
\end{eqnarray}\\
Now,  from Eq. (1), the density of states,
\begin{eqnarray}
\rho(\epsilon)=B\frac{\Gamma(\frac{d}{l} + 1)}{\Gamma(\frac{d}{l} + \sum_i \frac{1}{n_i}) }\epsilon^{\frac{d}{l}+\sum_i \frac{1}{n_i}-1} 
\end{eqnarray}
where, 
\begin{eqnarray}
B=\frac{gV_d C_d}{h^d a^{d/s}}\prod_{i=1} ^d \frac{\Gamma(\frac{1}{n_i} + 1)}{c_i ^{\frac{1}{n_i}}}
 \end{eqnarray}\\
Here,  $C_d=\frac{\pi^{\frac{d}{2}}}{\Gamma(d/2 + 1)} $, $g$ is the spin degeneracy factor,
$V_d=2^d\prod_{i=1} ^d a_i$ is the volume 
of an $d$-dimensional rectangular
whose $i$-th side has length $2a_i$ and  $\Gamma(l)=\int_0 ^\infty dx x^{l-1}e^{-x}$ is the gamma function.\\\\
The grand potential for the Bose system, 
\begin{eqnarray}
q=-\sum_\epsilon ln(1-zexp(-\beta \epsilon)) 
\end{eqnarray}
$\beta=\frac{1}{kT}$,
where $k$ being the Boltzmann Constant and $z=\exp(\beta \mu)$ is the fugacity, where $\mu$ being the chemical potential.
In  experiments
with trapped gases, thermal energies far exceed the level spacing\cite{anderson}. So, using the Thomas-Fermi semiclassical 
approximation\cite{thomas} and  re-writing the previous equation,
\begin{eqnarray}
 q=q_0 -\int_0 ^\infty \rho(\epsilon) ln(1-zexp(-\beta \epsilon))
\end{eqnarray}
So, using the density of states of Eq. (3) we finally get the grand potential,
\begin{eqnarray}
q=q_0+ B \Gamma(\frac{d}{l}+1)(kT)^\chi g_{\chi+1} (z)
\end{eqnarray}
where, $q_0=-ln(1-z)$, $\chi= \frac{d}{l} + \sum_{i=1} ^d \frac{1}{n_i}$
and  $g_l(z)$ is the Bose function.
Where,
\begin{equation}
 g_l(z)=\int_0 ^\infty \frac{x^{l-1}}{z^{-1}e^x-1}=\sum_{j=1} ^\infty\frac{z^l}{j^l}
\end{equation}
Point to note when $l>1$, as $z\longrightarrow 1$ (in condensed phase),
                \begin{eqnarray}
                g_l(1) = \sum_{r=1} ^\infty \frac{1}{r^l} = \zeta(l) \nonumber 
               \end{eqnarray}

\section{Thermodynamics of bose gas under generic power law potential in d dimension}
\subsection{Number of Particle}
The number of particles $N$ can be obtained,
\begin{eqnarray}
 N=z(\frac{\partial q}{\partial z})_{\beta,V}= N_0 + \frac{gC_n\Gamma(\frac{d}{l} +1) V_d \prod_{i=1} ^d  \Gamma(\frac{1}{n_i}+1 ) }{h^d b^{d/l} \prod_{i=1} ^d c_i ^{1/n_i} } (kT)^\chi g_\chi(z)
\end{eqnarray}
Here, $N_0=\frac{z}{1-z}$ is the ground state occupation number. \\ \\ 
Now, defining, \begin{eqnarray}
                V_d ' &=& V_d \prod_{i=1} ^d (\frac{kT}{c_i})^{1/n_i}\Gamma(\frac{1}{n_i} + 1)\\
                \lambda'&=& \frac{h b^{\frac{1}{l} }}{\pi ^{\frac{1}{2}} (kT) ^{\frac{1}{l}}} [\frac{d/2+1}{d/l+1}]^{1/d}
               \end{eqnarray}\\\\
It is noteworthy,
\begin{eqnarray}
&&\lim_{n_i\to\infty} V_d'=V_d  \\
&&\lim_{n_i\to\infty} \chi =\frac{d}{l}\\
&&\lim_{l\to 2, b\to \frac{1}{2m}} \lambda' =\lambda_0=\frac{h}{(2\pi mk T)^{1/2}}
\end{eqnarray}\\
So, if we choose $l=2$ and $b=\frac{1}{2m}$ from Eq. (14) 
we get $\lambda_0=\frac{h}{(2\pi mk T)^{1/2}}$, which is the thermal 
wavelength of nonrelativistic free massive boson. However, it should be noted that, 
when $l\neq 2$,  $\lambda'$ 
then depends on dimension. With $d=3$ and $d=2$, thermal wavelength of photons are respectively $\frac{hc}{2\pi^{1/2} kT}$ 
and $\frac{hc}{(2\pi)^{1/2}kT}$ which can be obtained from from Eq. (11) by choosing $b=c$, where $c$ being
the velocity of light. So, 
from the definition of $\lambda '$ with more general energy
spectrum, one can reproduce the thermal wavelength of of both massive and massless boson.\\ \\
The number of particle equation is then written as,
\begin{equation}
N-N_0= g \frac{V_d '}{{\lambda '} ^d}g_\chi(z)
\end{equation}
With $l=2$, $a=\frac{1}{2m}$, all $n_i\longrightarrow \infty$ the number of particle equation for free
massive boson in $d$ dimesnsional space can be 
obtained from Eq. (12), (13), (14), (15), 
\begin{equation}
 N-N_0= g \frac{V_d}{{\lambda_0} ^d}g_\frac{d}{2}(z)
\end{equation}
which is exactly like Ziff\cite{ziff}, which gives the exact equation for
number of particles at $d=3$\cite{pathria,huang}.\\\\
Now turning our concentration to find the critical temperature,
we know $T\longrightarrow T_c$, $\mu\longrightarrow 0$ or $z\longrightarrow1$. So, from equation (9), the 
critical density is,
\begin{eqnarray}
  N_c=  \frac{gC_n\Gamma(\frac{d}{l} +1)  \prod_{i=1} ^d  \Gamma(\frac{1}{n_i}+1 ) }{h^d b^{d/l} \prod_{i=1} ^d c_i ^{1/n_i} } (kT)^\chi \zeta(\chi)
\end{eqnarray}\\
And the critical temperature,
\begin{equation}
 T_c=\frac{1}{k}[\frac{N_c h^d b^{d/l}  \prod_{i=1} ^d c_i^{1/n_i} }{gC_n\Gamma(\frac{d}{l} +1)
 V_d \prod_{i=1} ^d  \Gamma(\frac{1}{n_i}+1 ) \zeta(\chi)}]^{\frac{1}{\chi} }
\end{equation}
Also from equation (17) and (18), we can obtain, the fraction of condensation,
\begin{eqnarray}
 \frac{N_0}{N}=1-(\frac{T}{T_c})^{^{\chi}}
\end{eqnarray}
For free massive boson, the two above equations become like below as in Ref. \cite{ziff,bhuiyan}
\begin{eqnarray}
T_c=(\frac{h}{\sqrt{2\pi mk^2}})^2 (\frac{N_c}{\zeta(d/2)})^{\frac{2}{d}}
\end{eqnarray}
\begin{equation}
 \frac{N_0}{N}=1-(\frac{T}{T_c})^{^{d/2}}
\end{equation}
From Eq. (20), it is seen if $d\longrightarrow \infty$, $T_c=(\frac{h}{\sqrt{2\pi m k}})^2$. From which we can say as $d$
approches infinity, the critical
temperature picks a non-zero value instead of being zero \cite{acharyya}. Also Eq. (20) and (21) exactly
reproduce the critical temperature and condensate fraction at $d=3$\cite{pathria,huang,C. J. PETHICK,ziff}.\\\\
Now, from equation (15), one can find the most general criterion for BEC to take place. 
It is seen that  
whether BEC to take place or not depends on $\chi$. Eq. (15) dictates BEC to take place when,
\begin{eqnarray}
\chi=\frac{d}{l}+\sum_i ^d \frac{1}{n_i}>1 
\end{eqnarray}
this is satisfied. This equation shows that the conditions for BEC to take place depend both on dimension $d$ as well
as on kinematics characteristic $l$ and exponent of power law $n_i$. When the potential is symmetric, i.e., 
$n_1=n_2=...=n_i=...n_d=n$, Eq. (22) becomes,
\begin{eqnarray}
 \frac{d}{l}+\frac{d}{n}>1
\end{eqnarray}
Now, putting $l=2$ for massive boson 
in the above equation we achieve the criterion for 
BEC to take place in case of symmetric potential which is same as Ref.\cite{sala}. 
\begin{eqnarray}
 \frac{d}{2}+\frac{d}{n}>1
\end{eqnarray}
In case of free massive boson, all $n_i\longrightarrow \infty$, so Eq. (24) becomes,
\begin{equation}
 \frac{d}{2}>1
\end{equation}\\
So, for free massive boson, we get the usual criterion that
$d$ should be greater than 2 for BEC to take place,
consistent with the result obtained in Ref.\cite{ziff,haug}.

\subsection{Internal Energy}
From the Grand Canonical Ensemble
internal energy $E$  is, \begin{eqnarray}
                             E&=&-(\frac{\partial q}{\partial \beta})_{z,V} \nonumber  \\
                              &=& \frac{gC_n\Gamma(\frac{d}{l} +1) V_d \prod_{i=1} ^d  \Gamma(\frac{1}{n_i}+1 ) }{h^d b^{d/l} \prod_{i=1} ^d c_i ^{1/n_i} } (kT)^{\chi+1} g_{\chi+1}(z)     
                              \end{eqnarray}
So, the internal energy below and greater than the critical temperature                             
  \begin{eqnarray}
   E = \left\{
     \begin{array}{lr}
       NkT \chi \frac{g_{\chi+1}(z)}{g_{\chi}(z)} &,  T>T_c\\
       (N-N_0) kT \chi \frac{\zeta(\chi+1)}{\zeta(\chi)}&,  T \leq T_c
     \end{array}
   \right.
\end{eqnarray}                       
In case of free massive bosons ($l=2$, $b=\frac{1}{2m}$ and $n_i\longrightarrow \infty$), 
one can find Eq. (27) reduces to,.
  \begin{eqnarray}
   E = \left\{
     \begin{array}{lr}
       NkT \frac{d}{2} \frac{g_{d/2+1}(z)}{g_{d/2}(z)} &,  T>T_c\\
       (N-N_0) kT \frac{d}{2} \frac{\zeta(d/2+1)}{\zeta(d/2)}&,  T \leq T_c
     \end{array}
   \right.
\end{eqnarray}                       
which is in accordance with Ziff\cite{ziff}, also produces the exact expression of $E$
for $d=3$\cite{pathria,huang}.\\\\
Now as $T>>T_c$, from Eq. (27) it is seen, the internal energy becomes, $E=NkT\chi$.
For free massive bosons it is $E=\frac{d}{2}NkT$, which becomes $\frac{3}{2}NkT$,
when $d=3$, thus $E$ approaches the classical value at high temperature.\\
\subsection{Entropy}
The entropy $S$ can be obtained from Grand Canonical Ensemble, 
\begin{eqnarray}
 S&=&kT(\frac{\partial q}{\partial T})_{z,V} -Nk\ln z +kq\nonumber  
 \end{eqnarray}
Again the entropy below and greater than the critical temperature
     \begin{eqnarray}
   S = \left\{
     \begin{array}{lr}
       N k [\frac{v_d'}{\lambda '^d}(\chi+1) {g_{\chi + 1}(z)}-\ln z] &,  T>T_c\\
       (N-N_0) k \frac{v_d'}{\lambda '^d}(\chi+1) \zeta(\chi+1)&,  T\leq T_c
     \end{array}
   \right.
\end{eqnarray}
From the above it is easy to see $S=0$ as $T\longrightarrow 0$, in accordance with the third law of thermodynamics.
As before, for free massive bosons ($l=2$, $b=\frac{1}{2m}$ and $n_i\longrightarrow \infty$), 
with the help of Eq. (12)-(14) one can find Eq. (29) reduces to,
\begin{eqnarray}
   S = \left\{
     \begin{array}{lr}
       N k [\frac{v_d}{\lambda ^d}(\frac{d}{2}+1) {g_{\frac{d}{2} + 1}(z)}-\ln z] &,  T>T_c\\
       (N-N_0) k \frac{v_d}{\lambda ^d}(\frac{d}{2}+1) \zeta(\frac{d}{2}+1)&,  T\leq T_c
     \end{array}
   \right.
\end{eqnarray}
which is in exactly like Ziff\cite{ziff}. Again at $d=3$ Eq. (30)
reduces to same expression for entropy as Ref.\cite{pathria,huang}\\
\subsection{Helmholtz Free Energy}
From the Grand Canonical Ensemble we get the expression of Helmholtz Free Energy,
\begin{eqnarray}
 A=-kTq+NkT\ln z
\end{eqnarray}
Now from the expression of grand potential we obtain the equation of Helmholtz Free Energy below and above $T_c$,
\begin{eqnarray}
   A =\left\{
     \begin{array}{lr}
       NkT \frac{g_{\chi+1} (z)}{g_{\chi} (z)} &,  T>T_c\\
       NkT \frac{\zeta({\chi+1})}{\zeta({\chi} )}&,  T\leq T_c
     \end{array}
   \right.
\end{eqnarray}         \\
In case of free massive boson the above expression reduces like below,
\begin{eqnarray}
   \frac{A}{NkT}=\left\{
     \begin{array}{lr}
       - \frac{g_{\frac{d}{2}+1} (z)}{g_{\frac{d}{2}} (z)}+\ln z &,  T>T_c\\
       - \frac{\zeta({\frac{d}{2}+1})}{\zeta({\frac{d}{2}} ) }  +\ln z &,  T\leq T_c
     \end{array}
   \right.
\end{eqnarray}         
exactly like Ziff\cite{ziff}. Now, for $d=3$, the above equation
produces the exact expression for Helmholtz free Energy\cite{pathria,huang}.
\subsection{Pressure}
Rewriting equation (9) stating the number of particles,
\begin{equation}
\frac{N-N_0}{V_d \prod_{i=1} ^d (\frac{kT}{c_i})^{1/n_i}\Gamma(\frac{1}{n_i} + 1)}=\frac{N-N_0}{V_d '}=\frac{g}{\lambda '^d}g_{\chi}(z)  \nonumber
\end{equation}\\
Now a very well known expression for the nonrelativistic $d-$dimensional ideal free Bose gas\cite{ziff},
\begin{equation}
\frac{N-N_0}{V_d}=\frac{g}{\lambda_0 ^d} g_{d/2}(z) \nonumber
\end{equation}\\
Comparing the above
equations, we can say $V_d '$ is a more generalized
extension of  $V_d$. It represents the effect of external potential on the performence of 
trapped bosons. Calling $V_d '$ the effective volume the grand potential can be rewritten as,
\begin{equation}
 q=q_0+g\frac{gV_d'}{\lambda '^d}g_{\chi+1}(z)
\end{equation}\\
So, the effective pressure \begin{eqnarray}
                            P'=\frac{1}{\beta}(\frac{\partial q}{\partial V_d '})=\left\{
     \begin{array}{lr}
       \frac{gkT}{\lambda '^d} g_{\chi+1} (z) &,  T>T_c\\
       \frac{gkT}{\lambda '^d} \zeta(\chi+1)  &,  T\leq T_c
     \end{array}
   \right.
                           \end{eqnarray}
Also can be rewritten as,
\begin{eqnarray}
                            P'=\frac{1}{\beta}(\frac{\partial q}{\partial V_d '})=\left\{
     \begin{array}{lr}
       \frac{NkT}{V_d '} \frac{g_{\chi+1} (z)}{g_{\chi}(z)} &,  T>T_c\\
       \frac{(N-N_0)}{V_d '} kT \frac{\zeta(\chi+1)}{\zeta (\chi)}  &,  T\leq T_c
     \end{array}
   \right.
                           \end{eqnarray}\\
The above equation is very general equation of state.
It is the equation of state for any dimensionality $d$, any dispersion relation 
of the form $(\propto$ $p^s$) having
any form of generic power law trap and obviously it is expected that it will reproduce the 
special case of free system. For free system the equation (36) becomes,
\begin{eqnarray}
                            P=\frac{1}{\beta}(\frac{\partial q}{\partial V_d })=\left\{
     \begin{array}{lr}
       \frac{NkT}{V_d } \frac{g_{d/2+1} (z)}{g_{\frac{d}{2}}(z)} &,  T>T_c\\
       \frac{(N-N_0)}{V_d } kT \frac{\zeta(\frac{d}{2}+1)}{\zeta (\frac{d}{2})}  &,  T\leq T_c
     \end{array}
   \right.
                           \end{eqnarray}
   which is in accordance with Ziff.                        
                           \\\\
                           Now, comparing Eq. (35) and (27) we get,
\begin{equation}
 P'V_d '=\frac{E}{\chi}
\end{equation}
For $d$-dimensional free Bose gas one can obtain from previous equation
\begin{equation}
 P V_d=\frac{2}{d}E
\end{equation}
This is an important and familiar relation, $P V=\frac{2}{3}E$ when $d=3$\cite{pathria,ziff,huang,C. J. PETHICK}.
This actually shows equation (38) is a very significant relation
for the Bose gas irrespective whether they are trapped or free.
And in case of trapped bosons effective volume and effective pressure
plays the same role as volume and pressure in current textbooks and literatures.\\\\
\subsection{Heat Capacity at Constant Volume}
Heat capacity at constant volume $C_v$ below and above $T_c$
\begin{eqnarray}
   C_V =T(\frac{\partial S}{\partial T})_{N, V} =\left\{
     \begin{array}{lr}
       N k [\chi(\chi+1)\frac{\nu'}{\lambda '^D}g_{\chi+1}(z)-\chi^2 \frac{g_{\chi}(z)}{g_{\chi-1}(z)}] &,  T>T_c\\
       N k \chi (\chi +1)\frac{\nu'}{\lambda '^D} \zeta(\chi+1)&,  T\leq T_c
     \end{array}
   \right.
\end{eqnarray}         \\\\
From this we can investigate  whether the specific heat will show a jump or not.
From Eq. (40)  we can obtain the difference between the heat capacities at constant volume, at $T_c$ as
\begin{eqnarray}
\Delta C_V\mid_{_{T=T_c}}=C_V\mid_{_{T_c^{-}}}- C_V\mid_{_{T_c^{+}}}=Nk\chi^2\frac{g_{\chi}(1)}{g_{\chi-1}(1)}
\end{eqnarray}\\
The above equation shows that,
the jump of heat capacity depends on $\chi$. 
$C_V$ will be discontinuous at $T=T_c$ if,
\begin{eqnarray}
\chi=\frac{d}{l}+\sum_{i=1} ^d \frac{1}{n_i}>2
\end{eqnarray}
is satisfied. And, $C_V$ will be continuous at $T=T_c$ if, $\chi$ satisfies
\begin{eqnarray}
1<\chi\leq2
\end{eqnarray}
So, in case of free massive boson (choosing $l=2$ and all $n_i\longrightarrow \infty$), $C_v$ will itself be discontinuous for $d>4$,
in agreement with Ziff\cite{ziff}.  And
in the high temperature limit of $C_v$ approches its classical value as it becomes $\chi Nk$ for trapped
system and $\frac{d}{2}Nk$ for free system, which is $\frac{3}{2}Nk$, when $d=3$. \\\\
\subsection{Isothermal Compressibilty}
The Isothermal Compressibilty of trapped Bose gas can be obtained,
\begin{eqnarray}
 \kappa_T=-V(\frac{\partial P'}{\partial V})_{_{N,T}}
\end{eqnarray}
Now using simple chain rule of partial derivative the above equation becomes
\begin{eqnarray}
 (\frac{\partial P'}{\partial V'})_{_{N,T}}=(\frac{\partial P'}{\partial z})_{_{N,T}}(\frac{\partial z}{\partial V'})_{_{N,T}}
\end{eqnarray}\\
Using equation (15) and (35), we get
\begin{eqnarray}
(\frac{\partial P}{\partial V})_{_{N,T}}=(\frac{-1}{NkT})\frac{g_{\chi-1}(z)}{g_\chi(z)}
\end{eqnarray}
Which concludes, 
\begin{eqnarray}
                            \kappa_T=\left\{
     \begin{array}{lr}
\frac{V}{NkT}\frac{g_{\chi-1}(z)}{g_\chi(z)}        &,  T>T_c\\
        \frac{V}{NkT}\frac{\zeta({\chi-1})}{\zeta(\chi)}   &,  T\leq T_c
     \end{array}
   \right.
                           \end{eqnarray}\\
which reproduces the same result for isothermal compressibilty of free massive Bose gas at $d=3$ \cite{pathria}.
And in the high temperature limit $\kappa_T$ takes the classical value for free system, which is $\frac{1}{P}$.\\ \\
\subsection{Heat Capacity at Constant Pressure}
Now, heat capacity at constant pressure $C_p$, 
\begin{eqnarray}
   C_P &=&T(\frac{\partial S}{\partial T})_{N, P} \nonumber\\
   &=&\left\{
     \begin{array}{lr}
 Nk  [   (\chi+1)^2 g_{\chi+1}^2(z)g_{\chi-1}(z)(\frac{\nu'}{\lambda '^D})^3-\chi(\chi+1)g_{\chi+1}(z)\frac{\nu'}{\lambda '^D}] &,T>T_c\\
  Nk [(\chi+1)^2 \zeta(\chi +1)\zeta(\chi-1)(z)(\frac{\nu'}{\lambda '^D})^3-\chi(\chi+1)\zeta(\chi+1)\frac{\nu'}{\lambda '^D}]  &,T\leq T_c
     \end{array}
   \right.
\end{eqnarray}         
In case of free massive Bose gas, the above equation reduces to,
\begin{eqnarray}
    C_P &=&T(\frac{\partial S}{\partial T})_{N, P} \nonumber\\
   &=&\left\{
     \begin{array}{lr}
  Nk [   (\frac{d}{2}+1)^2 g_{\frac{d}{2}+1}^2(z)g_{\frac{d}{2}-1}(z)(\frac{\nu'}{\lambda '^D})^3-\frac{d}{2}(\frac{d}{2}+1)g_{\frac{d}{2}+1}(z)\frac{\nu'}{\lambda '^D}] &,T>T_c\\
  Nk [(\frac{d}{2}+1)^2 \zeta(\frac{d}{2} +1)\zeta(\frac{d}{2}-1)(z)(\frac{\nu'}{\lambda '^D})^3-\frac{d}{2}(\frac{d}{2}+1)\zeta(\frac{d}{2}+1)\frac{\nu'}{\lambda '^D}]  &,T\leq T_c
     \end{array}
   \right.
\end{eqnarray}\\
For $d=3$, it coincides exactly with Ref. \cite{pathria,C. J. PETHICK} which diverges as $\frac{1}{T-T_c}$ when
the temperature approaches $
T_c$ from above. Again in the high 
temperature limit $C_p$ becomes $(\chi+1)Nk$ for trapped
system and $(\frac{d}{2}+1)Nk$ for free system, which is $\frac{5}{2}Nk$, when $d=3$. So, in 
the high temperature limit  $C_p$ approches its classical value.\\\\
Now, the ratio, $\gamma=(\frac{C_P}{C_V})$ for $T>T_c$ is given by,  
\begin{eqnarray}
  \gamma=\frac{(\chi+1)^2 \frac{g_{\chi+1}^2(z)g_{\chi-1}(z)}{g^3 _\chi (z)}-\chi(\chi+1)\frac{g_{\chi+1}(z)}{g_\chi(z)} } 
  {       \chi(\chi+1)\frac{\nu'}{\lambda '^D}g_{\chi+1}(z)-\chi^2 \frac{g_{\chi}(z)}{g_{\chi-1}(z)}}
\end{eqnarray}\\
Now, for $T>>T_C$, the above equation becomes,
\begin{eqnarray}
 \gamma=\frac{(\chi +1)^2-\chi(\chi +1)}{\chi(\chi +1)-\chi^2}=1+\frac{1}{\chi}
\end{eqnarray}
In case of free system, choosing all $n_i\longrightarrow \infty$, we get from the above equation,
\begin{equation}
\gamma=1+\frac{l}{d} 
\end{equation}
So, $\gamma$ equals $\frac{5}{3}$, when $d=3$ and $l=2$, thus obtaining the  classical value at high temperature limit.\\
\subsection{Clapaeyron Equation}
In any first order phase transition pressure is governed by Clapaeyron equation, in the transition line\cite{huang,ziff}. 
Like the BEC for free Bose gas at $d=3$, BEC for trapped Bose gas will also be a first order phase transition
if they obey the Clapaeyron equation. The Clapaeyron equation derived from Maxwell's relations is
\begin{eqnarray}
 \frac{dP}{dT}=\frac{\Delta s}{\Delta v}=\frac{l}{T\Delta v}
\end{eqnarray}
where, $l$, $\Delta s$ and $\Delta v$ are the latent heat, change in entropy and change in volume respectively. 
The effective  pressure in phase transition line is,
\begin{equation}
 P_0(T)=\frac{kT}{\lambda'^d}g_{\chi +1}(1)
\end{equation}
Now, differentiating with respect to $T$ leads, 
\begin{equation}
 \frac{dP_0}{dT}=\frac{1}{Tv_g}[(\chi+1)kT\frac{g_{\chi+1}(1)}{g_{\chi}(1)}]
\end{equation}\\
Now when two phases coexist
the non condensed  phase has specific volume $v_g$ whether the condensed phase 
has specific volume has specific 
volume  $0$, which concludes $\Delta v=v_g$. Which concludes the above equation to be Clapaeyron
equation where the latent heat of transition per particle in case of trapped boson is
\begin{eqnarray}
 L=\frac{g_{\chi +1}(1)}{g_{\chi }(1)}(\chi+1)kT
\end{eqnarray}\\
Therefore, BEC for trapped Bosons are also a first order phase transition in arbitrary dimensions. 
For free massive boson the latent heat per partcle becomes,
\begin{equation}
 L=\frac{g_{5/2}(1)}{g_{3/2 }(1)}(\frac{5}{2})kT
\end{equation}
agrees exactly with Ref.\cite{huang}.
\section{Discussion}
Properties of ideal Bose gas in the presence of an external generic power law potential  (not essentially symmetric) 
are discussed in this section.
The study  done by  
density of states approach in $d$ dimensional space  allows us to investigate the two phase system that 
is both in the condensed and non condensed phase across the phase transition point.  The most general 
conditions for BEC as well as the continuity conditions of specific heat for ideal Bose gas trapped in 
generic power law potential are acquired with any kinematics characteristic in $d$ dimensional space.\\ \\
At first, the density of states and grand potential of Bose gas under generic power law potential in d dimension
has been calculated from which we can easily derive the expression for number of particle, which gives us
the first insight about the condensed phase. Using number of particle equation, the general
criterion for BEC has been obtained from which, one can achieve the same conclusion
for symmetric potential\cite{sala}, harmonic potential\cite{dal} as well as more simplified system of 
free massive boson\cite{pathria,huang,ziff}. The fraction of condensate as well as the 
critical temperature was achieved in this process, predicting the exact equation for 
free
system\cite{ziff}. Now turning our attention in thermodynamic
quantities such as, internal energy $E$, entropy $S$, free energy $A$ all of which 
are evaluated from the grand potential.
The obtained expressions coincides with available literature when the potential is symmetric
($n_1=n_2=..=n_i=..=n_d$)\cite{sala}, harmonic ($n_1=n_2=..=n_i=..=n_d=2$)\cite{dal} 
or the system is free (all $n_i\longrightarrow \infty$)\cite{ziff}. 
Now, from the pressure equation we have derived the equation of state which determines a relation
among the thermodynamic generalized coordinate $y$ and the thermodynamic generalized force $Y$, and temperature
$T$ of the system. Point
to note in 
case of trapped boson we have seen, pressure $P$ is replaced by effective pressure $P'$ and volume $V$ is replaced by
effective volume $V'$. It is seen that the high temperature limit of Bose gas is just classical system obeying 
Maxwell-Boltzmann distribution. Therefore, the idea of effective volume and pressure is not  only  valid for trapped 
Bose gas but also for trapped
classical system as well trapped  Fermi gas. For free system, both effective pressure and effective volume
reduces to pressure and volume. So we can say effective volume and pressure to be  more general notion for trapped system, 
which enables us to treat the trapped atomic gases as free one.  The other derived quatities for trapped system
such as specific heat at constant volume $C_v$, 
specific heat at constant pressure $C_p$, isothermal compressibilty $\kappa$ all can reproduce their familiar form found 
in literature\cite{pathria, huang, C. J. PETHICK, ziff} in case of $d$ dimensional system\cite{ C. J. PETHICK, ziff} 
as well as in $d=3$\cite{pathria,huang}.\\\\
Now, lets turn our focus towards general criterion of BEC (Eq. 22) and jump of $C_v$ at $T=T_c$ (Eq. 42). 
In general it is seen that there is no first order\cite{pathria,huang,ziff}
as well as no second order \cite{Hohenberg} phase transition for $d\leqslant 2$ for free system. 
Eq. (25) derived from the general criterion Eq. (22) also suggests the same for the free ideal Bose gas.
Now turning our attention to trapped system,  consider 
a one dimensional system with $n_1=2$ and $l=2$. Eq. (22) then dictates, no BEC for such system as $\chi<1$.
In case of two dimensional Bose gas ($d=2$), with $n_1=n_2=2$ (i.e. harmonic potential) we can find it fulfills 
Eq. (22), so BEC exists in such system. That means, we have seen the case 
when BEC can exist for $d\leqslant 2$, with appropiate
choice of power law exponent,
which was not the case for ideal free Bose gas. In case of jump of $C_v$ in free Bose system, it is seen that,  
there is a jump of $C_v$ at $T=T_c$,
when $d>4$\cite{ziff}. For instance, taking $d=3$, $l=2$ and $n_i\longrightarrow \infty$ we see that there is no discontinuity 
at $d=3$ for free system as $\chi<2$. But again one can obtain jump in $C_v$ at $T=T_c$,
for trapped system  when $d\leqslant 4$ with appropiate choice of $n_i$. For example, 
taking $d=3$ and $l=2$, with $n_1=n_2=n_3=2$ one
can obtain from Eq. (42) that, 
$\chi>2$. So, in this case we can obtain a jump at $d=3$  in $C_v$ at $T=T_c$ for trapped
system, although not a property for free Bose gas. 
Neveretheless, Eqs. (40) and (41) shows that
although $C_{T_c ^+}$ is not a monotonical function
of $\chi$, both $C_{T_c ^-}$ and $\Delta C_{T_c }$ increase monotonically
as the parameter $\chi$ increases. This indicates that, in general, in the vicinity of $T_c$, larger the value of $\chi$,
larger the energy needed to raise the Bose system to higher temperature state, partticularly when the 
system is in condensed phase. On the other hand, when the temperature is very low, $C_{v(T<T_c)}$ decreases as $\chi$
increases and the larger the 
value of $\chi$, more quickly $C_{v(T<T_c)}$ tends to zero. Also larger value of $\chi$ indicates
larger fraction of bosons to be in the ground state, such that less number of bosons excited when temperature raises.\\\\
The Clapaeyron equation applies to any first order phase transition.
From the expression of effective pressure it is shown that 
trapped bosons in arbitrary dimension does satisfies the 
Clapaeyron equation, from which we can say BEC is a first order phase transition
with or without the existense of external potential. The latent heat of transition per particle in case of trapped system 
is also obtained from this equation, which reduces to exact expression of laten heat in case of free massive bosons at $d=3$\cite{huang}.

\section{Conclusion}
From
complete thermodynamics   
of ideal Bose gas trapped in generic power law potential we have derived the general criterion for BEC. Also,
the calculated physical quantities reduce to expressions available in the literature, 
with appropiate choice of power law exponent.
In
this manuscript, we have restricted our discussion in case of ideal system. It will be vey interesting 
to see the effect of interaction on general criterion of BEC.
\section{Acknowledgement}
I would like to thank Fatema Farjana, for her efforts to help me present this work and Mishkat Al Alvi 
for showing the typographic mistakes.

  \end{document}